\documentclass[prd,superscriptaddress]{revtex4}

\usepackage{amsmath}
\usepackage{bm}
\usepackage{graphics}
\usepackage{color}

\def\gsim{\;\rlap{\lower 2.5pt
 \hbox{$\sim$}}\raise 1.5pt\hbox{$>$}\;}
\def\lsim{\;\rlap{\lower 2.5pt

   \hbox{$\sim$}}\raise 1.5pt\hbox{$<$}\;}

\def\be{\begin{equation}}
\def\ee{\end{equation}}
\def\ba{\begin{eqnarray}}
\def\ea{\end{eqnarray}}

\def\k0{k_0^{\rm{p}}}

\begin{document}

\title{Pulsar Timing Residuals Induced by  Gravitational Waves from  Single Non-evolving  Supermassive Black Hole Binaries with Elliptical Orbits}

\author{Ming-Lei Tong}

\thanks{Email: mltong@ntsc.ac.cn}

\affiliation{National Time Service Center, Chinese Academy of Sciences, Xi'an, Shaanxi 710600,  China }
\affiliation{Key Laboratory of Time and Frequency Primary Standards, Chinese Academy of Sciences, Xi'an, Shaanxi 710600,  China }

\author{Bao-Rong Yan}

\affiliation{National Time Service Center, Chinese Academy of Sciences, Xi'an, Shaanxi 710600,  China }
\affiliation{Key Laboratory of Precision Navigation and Timing Technology, Chinese Academy of Sciences, Xi'an, Shaanxi 710600,  China }

\author{Cheng-Shi Zhao}

\affiliation{National Time Service Center, Chinese Academy of Sciences, Xi'an, Shaanxi 710600,  China }
\affiliation{Key Laboratory of Time and Frequency Primary Standards, Chinese Academy of Sciences, Xi'an, Shaanxi 710600,  China }

\author{Dong-Shang Yin}

\affiliation{National Time Service Center, Chinese Academy of Sciences, Xi'an, Shaanxi 710600,  China }
\affiliation{Key Laboratory of Time and Frequency Primary Standards, Chinese Academy of Sciences, Xi'an, Shaanxi 710600,  China }

\author{Shu-Hong Zhao}

\affiliation{National Time Service Center, Chinese Academy of Sciences, Xi'an, Shaanxi 710600,  China }
\affiliation{Key Laboratory of Time and Frequency Primary Standards, Chinese Academy of Sciences, Xi'an, Shaanxi 710600,  China }

\author{Ting-Gao Yang}

\affiliation{National Time Service Center, Chinese Academy of Sciences, Xi'an, Shaanxi 710600,  China }
\affiliation{Key Laboratory of Time and Frequency Primary Standards, Chinese Academy of Sciences, Xi'an, Shaanxi 710600,  China }

\author{Yu-Ping Gao}

\affiliation{National Time Service Center, Chinese Academy of Sciences, Xi'an, Shaanxi 710600,  China }
\affiliation{Key Laboratory of Time and Frequency Primary Standards, Chinese Academy of Sciences, Xi'an, Shaanxi 710600,  China }

\begin{abstract}
 The pulsar timing residuals induced by gravitational waves from  non-evolving single binary
 sources with  general elliptical orbits will be analyzed. For different orbital
eccentricities, the timing residuals present different properties. The standard deviations
of the timing residuals induced by a fixed gravitational wave source  will be calculated for different values of the eccentricity. We will also analyze the timing residuals of PSR J0437-4715 induced by one of the best known single gravitational wave sources, the  supermassive black hole binary in the blazar OJ287.

\


\end{abstract}

\maketitle

\large

The existence of gravitational waves (GWs) was proposed just after
the publication of general relativity. Hereafter, the detection of GWs has been a very
interesting field to which are payed much attention by many scientists.
The sources of GWs are diverse, but can be divided into three types.
The first  is the continuous sources$^{[1]}$, including the inspiral of binary compact stars,
the  rotating isolated neutron stars, and the neutron stars in the x-ray binary star systems.
The second is the instantaneous  sources$^{[2]}$, such as supernovae and coalescing binary systems.
The third is the stochastic gravitational wave background, including relic gravitational waves generated in the early universe$^{[3-6]}$, random signals from an ensemble of independent
binary star systems$^{[7,8]}$, and the GWs produced by oscillating cosmic string loops$^{[9,10]}$.
The first indirect evidence for
the existence of GW emission was provided by the observations of the binary pulsar
B1913+16$^{[11]}$. On the other hand, many methods of direct detection of GWs have
 been proposed and tried for a long time, even though there is no assured result so far.
Various GW detectors were constructed or proposed, such as ground-based $^{[12]}$ and space-based$^{[13]}$ interferometers, pulsar timing arrays$^{[14-17]}$,  waveguide$^{[18,19]}$, Gaussian beam$^{[20,21]}$, and even the anisotropies
and polarizations of the cosmic microwave background radiation$^{[22-24]}$, aiming at different detection frequencies.

Recently, more and more pulsars are founded, and, moreover, the measurement technique is more and more precise with the improvement of radio telescopes. These make
 pulsar timing arrays (PTAs) be powerful in detecting  GWs directly.  Due to the existence of GWs
 passing through the path between the pulsars and the earth, the times-of-arrival (TOAs) of the pulses radiated from pulsars will be fluctuated. As shown in Ref.[25], a stochastic GW background can be detected by searching for correlations in the timing residuals of an array of millisecond pulsars spread over the sky. On the other hand, single sources of GWs are also important in the observations of pulsar timing array$^{[26]}$ or an individual pulsar$^{[27]}$.
 Currently,
 there are several  PTAs running,
such as the Parkes Pulsar Timing Array (PPTA)$^{[28]}$ ,
European Pulsar Timing Array (EPTA)$^{[29]}$,
the North American Nanohertz Observatory for Gravitational Waves (NANOGrav)$^{[30]}$,
and the International Pulsar Timing Array (IPTA)$^{[31]}$ formed by the aforesaid three PTAs.
Moreover, a much more sensitive Square Kilometer Array (SKA)$^{[32]}$
is also under planning.

 The typical response frequencies of a PTA to GWs lie in the range of $\sim10^{-9}-10^{-7}$ Hz, where the lower frequency limit is the inverse of  observation time span and the upper frequency limit corresponding to the observation time interval (e.g., two weeks).
In this range of frequencies,
 the gravitational radiation by supermassive black hole binaries (SMBHBs)$^{[7,33]}$
may be the  major  targets of PTAs.
  As analyzed in Ref.[26], a PTA is sensitive to  the nano-hertz GWs from SMBHB systems with masses of $\sim 10^8-10^{10}M_\odot$ less than $\sim10^5$ years
 before the final merger. Binaries with more than $\sim10^3-10^4$ years before merger can be
 treated as  non-evolving GW sources. The non-evolving SMBHBs are believed to be the dominant population, since they have lower masses and longer rest lifetimes.

In this letter, we  analyze  theoretically the timing residuals of an individual pulsar induced by single  non-evolving GW sources of SMBHBs. This means that the time before the final merge would be $10^5-10^6$ years. Larger rest lifetimes of SMBHBs will not lie in the detecting range of PTAs $^{[26]}$. In most literatures, the orbit of a SMBHB is
 often assumed to be circular, i.e., the eccentricity $e=0$. This may be reasonable since $e$ is
 decaying along with time due to the evolution of the elliptic orbit under back-reaction of the binary system$^{[34]}$, especially at
 the late time before merge. However, if the SMBHBs are in the phase long enough before merge, the eccentricity may be not zero. For example,  one of the best-known candidates for a SMBHB system emitting
GWs with frequency detectable by pulsar timing is in the blazar OJ287$^{[35]}$, with an orbital
eccentricity $e\sim0.7$$^{[36,37]}$. The gravitational
radiation from the SMBHB in the blazar OJ287 is very strong with the amplitude $h\sim10^{-15}$ as we shall see, which will be  an appropriate detecting target of pulsar timing arrays.       To give a general analysis, we will assume the motion
 of the sources is on an elliptic Keplerian orbit. Throughout this paper, we use units in which the light speed $c=1$.

 The usual equation for the
 relative orbit ellipse is $^{[38]}$
 \be
 r=\frac{a(1-e^2)}{1+e \cos(\theta-\theta_p)},
 \ee
 where $r$ is the relative separation of the binary components, $a$ is the semi-major axis, and $\theta_p$ is the value
 of $\theta$ at the periastron. The binary period is given by
 \be
 P=\left(\frac{4\pi^2a^3}{GM}\right)^{1/2},
 \ee
 where $G$ is the gravitational constant, and $M$ is the total mass of the binary system.
 The differential equation for the Keplerian motion can be written as
 \be
 \dot{\theta}=\frac{2\pi}{P}(1-e^2)^{-3/2}[1+e \cos(\theta-\theta_p)]^2.
  \ee
Since different values of $\theta_p$ stand for  different choices of the ``zero point'' of $\theta$, and in turn correspond to different choices of the starting point of  time, if one
chooses the integration constant in Eq. (3) so that $t=0$ when $\theta=0$.   For a concrete discussion, we fix the value of $\theta_p=180^\circ$ without losing generality in the following. Integrating Eq. (3), one has
\ba\nonumber
&&\frac{2\pi}{P}t(\theta)=(2n+1)\pi-\arccos\left(\frac{e-\cos\theta}{1-e\cos\theta}\right)
+\frac{e(1-e^2)^{1/2}\sin \theta}{1-e\cos\theta}, \  2n\pi\leq\theta\leq(2n+1)\pi;\\
&&\frac{2\pi}{P}t(\theta)=(2n+1)\pi+\arccos\left(\frac{e-\cos\theta}{1-e\cos\theta}\right)
+\frac{e(1-e^2)^{1/2}\sin \theta}{1-e\cos\theta}, \  (2n+1)\pi\leq\theta\leq(2n+2)\pi,
 \ea
where $n$ stands for integers.  In Fig.1, we show the properties of $t$ as a function of $\theta$ with different values of $e$.

 \begin{figure}
\resizebox{90mm}{!}{\includegraphics{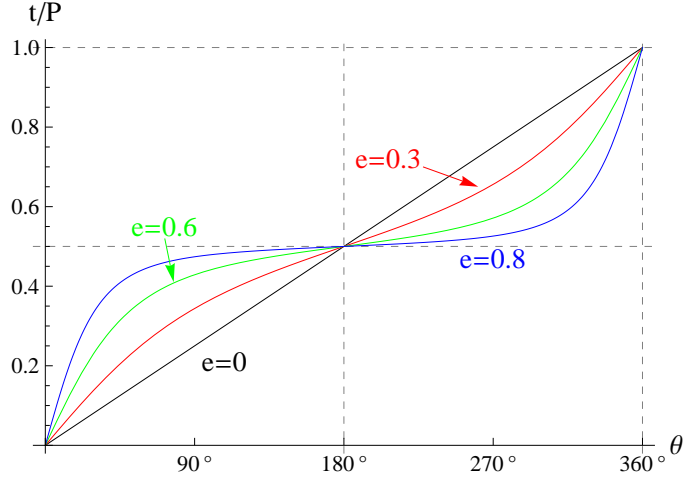}
}
\caption{\label{betan}
The time $t$ as a function of $\theta$ for different values of $e$.
}
\end{figure}

Let $\{\hat{i},\hat{j},\hat{k}\}$ be the fundamental celestial
frame where the origin locates at the Solar System Barycenter (SSB).
Then the unit vector of a GW source is
\be
\hat{d}=\cos\delta(\cos\alpha \hat{i}+\sin\alpha \hat{j})+\sin\delta\hat{k},
\ee
where $\alpha$ and $\delta$ are the right ascension and declination of the binary source, respectively. Define orthonormal vectors on the celestial sphere by$^{[38]}$
\ba\nonumber
&&\hat{\alpha}\equiv-\sin\alpha\hat{i}+\cos\alpha\hat{j},\\
&&\hat{\delta}\equiv-\sin\delta(\cos\alpha\hat{i}+\sin\alpha\hat{j})+\cos\delta\hat{k}.
\ea
The waveform of GWs in the transverse-traceless (TT) gauge  can generally be written as
\be
h_{ab}^{\rm{TT}}(t,\hat{\Omega})=h_{+}(t)\epsilon_{ab}^{+}(\hat{\Omega})
+h_\times(t)\epsilon_{ab}^\times(\hat{\Omega}),
\ee
where $\hat{\Omega}=-\hat{d}$ is the unit vector pointing from the GW source to the SSB.
For  a SMBHB
with a elliptical orbit, the polarization amplitudes of the emitting GWs are$^{[38]}$
\ba
&&h_+(\theta)=H\{{\cos(2\phi)}[A_0+eA_1+e^2A_2]
-\sin(2\phi)[B_0+eB_1+e^2B_2]\}, \\
&&h_\times(\theta)=H\{{\sin(2\phi)}[A_0+eA_1+e^2A_2]
+\cos(2\phi)[B_0+eB_1+e^2B_2]\},
\ea
with $\phi$ being the orientation of the line of nodes, which is defined to be the intersection
of the orbital plane with the tangent plane of the sky. The parameters in Eqs. (8) and (9) are
\ba\nonumber
&&H\equiv\frac{4G^2m_1m_2}{a(1-e^2)d},\\ \nonumber
&&A_0=-\frac{1}{2}[1+\cos^2(\iota)]\cos(2\theta),\\\nonumber
&&B_0=-\cos(\iota)\sin(2\theta),\\\nonumber
&&A_1=-\frac{1}{4}\sin^2(\iota)\cos\theta+\frac{1}{8}[1+\cos^2(\iota)][5\cos\theta+\cos(3\theta)],\\\nonumber
&&B_1=\frac{1}{4}\cos(\iota)[5\sin(\theta)+\sin(3\theta)],\\\nonumber
&&A_2=\frac{1}{4}\sin^2(\iota)-\frac{1}{4}[1+\cos^2(\iota)],\\
&&B_2=0,
\ea
where we have chosen $\theta_n=0$, the value of $\theta$ at the line of nodes. $m_1$ and $m_2$ are the masses of the two components of the binary system,  $d$ is the comoving distance from the binary system to the SSB,
 and  $\iota$ is the angle of inclination of the orbital plane to the tangent plane of the sky. The polarization tensors are$^{[26,38,39]}$
 \be
 \epsilon_{ab}^+(\hat{\Omega})=\hat{\alpha}_a\hat{\alpha}_b-\hat{\delta}_a\hat{\delta}_b,
 \ee
 \be
 \epsilon_{ab}^\times(\hat{\Omega})=\hat{\alpha}_a\hat{\delta}_b-\hat{\delta}_a\hat{\alpha}_b,
 \ee


The GW will cause a fractional shift in frequency, $\nu$, that can be defined by a redshift $^{[15,39,40]}$
\be
z(t,\hat{\Omega})\equiv\frac{\delta\nu(t,\hat{\Omega})}{\nu}=-\frac{1}{2}\frac{\hat{n}^a\hat{n}^b}{1+\hat{n}\cdot
\hat{\Omega}}\epsilon_{ab}^{A}(\hat{\Omega})\Delta h_A(t),
\ee
where
\be
\Delta h_A(t)=h_A(t_e)-h_A(t_p).
\ee
Here $A$ denotes ``$+,\times$'' and  the standard Einstein summing convention was used.  $t_e$
and $t_p$ are the time at which the GW passes the earth and pulsar, respectively.
Henceforth, we will drop the subscript ``$e$'' denoting the earth time unless otherwise noted.
 The unit vector, $\hat{n}$, pointing from the SSB to the pulsar is  explicitly written as
 \be
 \hat{n}=\cos{\delta_p}[\cos\alpha_p\hat{i}+\sin\alpha_p\hat{j}]+\sin{\delta_p}\hat{k},
 \ee
 where    $\alpha_p$ and $\delta_p$ are the right ascension and declination of the pulsar, respectively.
From geometry on has$^{[39,40]}$
 \be
 t_p=t-D_p(1-\cos\eta),
 \ee
 where $D_p$ is the distance to the pulsar and $\cos\eta=-\hat{n}\cdot\hat{\Omega}=\hat{n}\cdot\hat{d}$ with
 $\eta$ being the angle between the pulsar direction and the GW source direction.
 Combining Eqs.(10)-(14), we obtain
 \ba
 z(t,\hat{\Omega})=-\frac{1}{2}(1+\cos\eta)\{\cos(2\lambda)
 [h_+(t)-h_+(t_p)]
 +\sin(2\lambda)[h_\times(t)-h_\times(t_p)]\},
 \ea
 where $\cos\eta=\sin\delta_p\sin\delta+\cos{\delta_p}\cos{\delta}\cos(\alpha-\alpha_p)$, and
 $\lambda$ is defined as$^{[38]}$
 \be
 \tan\lambda\equiv\frac{\hat{n}\cdot\hat{\delta}}{\hat{n}\cdot\hat{\alpha}}=
 \frac{\cos\delta_p\sin\delta\cos(\alpha-\alpha_p)-\sin\delta_p\cos\delta}{\cos\delta_p
 \sin(\alpha-\alpha_p)},
 \ee
 where Eqs. (6) and (15) were used. It can be found from Eq. (17) that,
 the frequency of the pulses from the pulsar will suffer no shift from the GW
 for $\eta=0^\circ$ and $\eta=180^\circ$ allowing for Eq. (16).

 The pulsar timing residuals induced by GWs can be computed by integrating the redshift given
 in Eq.(17) over the observer's local time$^{[17,26,39,40]}$:
 \be
 R(t,\hat{\Omega})=\int_0^tz(t',\hat{\Omega})dt'.
 \ee
In the following, we will take PSR J0437-4715 as an example to calculate the time residuals $R$
induced by single GW sources.  We take the distance of PSR J0437-4715 to be $D_p\sim157$ pc$^{[41]}$. As we focus on the properties of $R$ due to different orbital eccentricities of the binary system in this letter, other parameters will be chosen to be fixed values. Concretely,
we choose $\phi=0$, $\iota=0$, $\lambda=22^\circ$, and  $\eta=120^\circ$. Since the detecting window of pulsar timing is $10^{-9}-10^{-7}$ Hz and the frequencies of the GWs from a SMBHB
with an elliptical orbit are a few or tens times of the orbital frequency of the binary system,
we set the orbital period to be $P=10^9s$. Furthermore, we assume $H=10^{-15}$, which is below the upper limit given by Ref.[42]. With the concrete values given above, one can calculate the timing residuals, which have been illustrated in Fig.2 for different values of $e$. One can see
that $R(t)$ is periodic due to the periodicity of the polarization amplitudes in Eq.(8) and (9). Note that, the time residual $R$ is proportional to the amplitude $H$.
\begin{figure}
\resizebox{90mm}{!}{\includegraphics{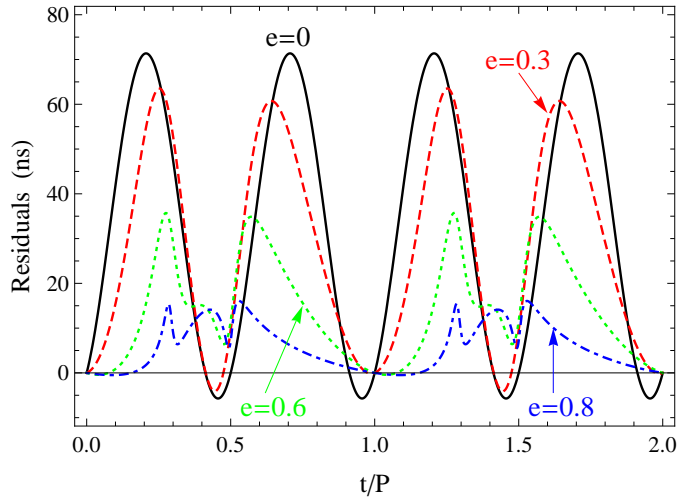}
}
\caption{\label{betan}
The time residuals of PSR J0437-4715 induced by a single GW source at a fixed position with
the eccentricities  $e=0$, $e=0.3$, $e=0.6$, and $e=0.8$, respectively, for two periods.
}
\end{figure}

It is also interesting to calculate the standard deviation of the time residuals, $\sigma_R$, which is defined as
\be
\sigma_R=\left[\frac{1}{T}\int_0^T R^2(t)dt-\left(\frac{1}{T}\int_0^T R(t)dt\right)^2\right]^{1/2},
\ee
where $T$ can be chosen as the period $P$, due to the periodicity of $R(t)$. Using the same
sets of parameters given above, the resulting values of $\sigma_R$ for different $e$ are listed
 in Table 1. One can see that, for the same sets of parameters, a larger $e$ leads to a smaller
 value of $\sigma_R$.

\vskip 2mm

\noindent{\footnotesize Table 1. The resulting  $\sigma_R$ for different values of $e$.

\vskip 2mm \tabcolsep 6pt

\centerline{\footnotesize
\begin{tabular}{c|cccc}\hline
$e$& $e=0$ & $e=0.3$ & $e=0.6$& $e=0.8$  \\
\hline $\sigma_R (ns)$
 & 27.3 & 22.4 & 11.2 & 5.2  \\
\hline
\end{tabular}}}

\vskip 0.5\baselineskip

Below, we analyze the time residuals of PSR J0437-4715 induced by the SMBHB in  the blazar OJ287.
The right ascension and declination of PSR J0437-4715 are $\alpha_p=69.3^\circ$ and
$\delta_p=-47.3^\circ$, respectively, given by Ref.$[41]$. On the other hand, the right ascension and declination of OJ287 are $\alpha=133.7^\circ$ and $\delta=20.1^\circ$, respectively, given by Ref.$[43]$. With the help of Eq.(18), one get $\lambda=52.3^\circ$. Combining Eqs.(5) and (15), one has $\eta=88.7^\circ$. Taking the values of parameters obtained in Ref.[37]: $m_1=1.84\times10^{10}M_\odot$, $m_2=1.46\times10^8M_\odot$, $P=3.76\times10^8\,s$, $e=0.67$,  $a=0.0535$ pc, and $z=0.306$ corresponding to $d\sim1311$ Mpc. Substituting these parameters into the first formula of Eq.(10), one obtain $H\simeq6.4\times10^{-16}$. 
 Before calculation using Eqs.(13) and (14), one should check that whether the SMBHB in the blazar OJ287 can be treated as a non-evolving binary system. Similar to the  analysis in Ref.[26], if the characteristic frequency difference between the pulsar and the Earth term is larger than the frequency resolution of PTAs $\sim10^{-9}$ Hz, one should consider the evolution effects of the binary system. This requires $\delta f\geq 10^{-9}$ Hz, where $\delta f\approx \dot{f}_0 D_p$ with the characteristic frequency $f_0=1/P$. From the dynamic evolution of the binary system, one has
\be
\dot{f}_0=\frac{3}{16\pi}\left(\frac{5}{256}\right)^{3/8}\frac{[f(e)]^{-3/8}}
{(GM_c)^{5/8}\tau^{11/8}(1+z)^{5/8}},
\ee
where $\tau$ is the time before the final merger of the binary system at the observer,  $M_c$ is the chirp mass of the binary system, and $f(e)$ is the correction due to the elliptic orbit. Employing $\tau=18000$ years given in Ref.[44], $f(0.67)\simeq20$, and the values of other related parameters given above, one can estimate the frequency difference to be $\delta f\approx10^{-9}$ Hz, which just reaches the frequency resolution of PTAs.
 For the consistency,  we will ignore the evolution effects of the binary in OJ287 for the moment, since
 we focus on the non-evolving GW sources in this paper.
  The  evolution effects of the binary in  OJ287 would be studied elsewhere separately.
Fig.3 presents the timing residuals for two periods. Here we also set $\phi=0$ and $\iota=0$. The resulting standard deviation is $\sigma_R\simeq2.3\,ns$. Note that this result is larger than that shown in Ref.$[17]$, where the authors claimed that the induced timing residuals due to OJ287 are significantly less than $1\,ns$.
However, there are two differences between them. Firstly, the values of the parameters were chosen differently in Ref.$[17]$, which leads to one order of magnitude lower in estimating the GW amplitude $H\simeq2.7\times10^{-17}$. Secondly, the authors assumed an evolving GW source during simulations.

\begin{figure}
\resizebox{90mm}{!}{\includegraphics{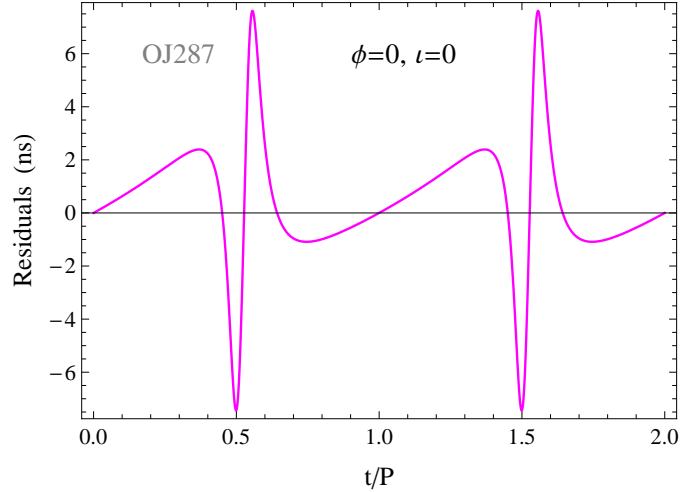}
}
\caption{\label{betan}
The time residuals of PSR J0437-4715 induced by a SMBHB in the blazar OJ287.
}
\end{figure}

In summary, we have shown the  properties of pulsar timing residuals
induced by a non-evolving  SMBHB with an elliptical orbit for different eccentricities.
It can be found that the forms of the timing residuals are quite different
for different $e$, and would also depend on other parameters such as
$\phi$, $\iota$, $\lambda$ and $\eta$, which could be studied in the subsequent
papers. For the same set of parameters, we found that a larger value of $e$ gives
a smaller standard deviation of the timing residuals, $\sigma_R$. In addition, we
calculated the timing residuals of PSR J0437-4715 induced by the SMBHB in the blazar
OJ287. The resulting standard deviation of the timing residuals is found to be
$\sigma_R=4.3\,ns$. Based on the analysis in this paper, similar calculations can be applied
to  other pulsars. On the other hand, one can also  study
the evolving  single GW sources.


\small
\begin{thebibliography}{88}

\bibitem{1} Peters P C 1964 {\it Phys. Rev.} {\bf136} 1224

\bibitem{2} Thorne K S and Braginskii V B 1976 {\it ApJ} {\bf 204} L1


\bibitem{3}  Grishchuk L P 1975
     {\it Sov. Phys. JETP} {\bf 40} 409

\bibitem{4}  Starobinsky A A 1979
        {\it JEPT Lett.} {\bf 30} 682

\bibitem{5} Zhang Y, Yuan, Y F, Zhao W  et al 2005
   {\it Class. Quant. Grav.} {\bf 22} 1383

\bibitem{6} Tong M  and Zhang Y  2009 {\it Phys. Rev.} D {\bf 80} 084022
054301

\bibitem{7} Jaffe A H and Backer D C 2003 {\it ApJ} {\bf583} 616

\bibitem{8} Wyithe J S B and Loeb A 2003 {\it ApJ} {\bf590}, 691

\bibitem{9} Vilenkin A 1982 {\it Phys. Lett. B} {\bf107} 47

\bibitem{10} Damour T and Cilenkin A 2005 {\it Phys. Rev.} D {\bf71} 063510
\bibitem{11} Hulse R A and Taylor J H 1974 {\it ApJ} {\bf 191} L59

\bibitem{12} http://www.ligo.caltech.edu/

\bibitem{13} http://elisa-ngo.org/

\bibitem{14} Sazhin M V 1978 {\it Soviet Astron.} {\bf 22} 36

\bibitem{15} Detweiler S 1979 {\it ApJ} {\bf234} 1100

\bibitem{16} Jenet F A, Hobbs G B, Lee K J, and Manchester R N 2005 {\it ApJ} {\bf625} L123

\bibitem{17} Hobbs G, Jenet F and Lee K J et al 2009 {\it Mon. Not. R. Astron. Soc.} {\bf 394}
1945

\bibitem{18}  Cruise A M 2000 {\it Class.Quant.Grav.} {\bf 17} 2525

\bibitem{19}  Tong M L and Zhang Y 2008 {\it
                    Chin. J. Astron. Astrophys.} {\bf 8} 314

\bibitem{20} Li F Y,  Tang  M X and  Shi D P 2003 {\it
           Phys. Rev.} D {\bf 67} 104008

\bibitem{21}  Tong M L,  Zhang Y   and  Li F Y 2008 {\it
         Phys. Rev.} D {\bf78} 024041

\bibitem{22}   Zaldarriaga M and   Seljak U, 1997 {\it
                     Phys. Rev.} D {\bf 55} 1830

\bibitem{23}  Kamionkowski M,  Kosowsky A and  Stebbins A 1997 {\it
                 Phys. Rev.} D {\bf 55} 7368

\bibitem{24}     Zhao W and   Zhang Y, 2006 {\it Phys. Rev.} D {\bf 74} 083006

\bibitem{25} Hellings R W and Downs G S 1983 {\it ApJ} {\bf265} L39

\bibitem{26} Lee K J, Wex N, Kramer M et al 2011 {\it Mon. Not. R. Astron. Soc.} {\bf414} 3251

\bibitem{27} Jenet F A, Lommen A, Larson S L, and Wen L  2004 {\it ApJ} {\bf606} 799

\bibitem{28} Manchester R N, Hobbs G, Bailes M et al 2013 {\it PASA} {\bf30} 17

\bibitem{29} van Haasteren R, Levin Y, Janssen G H et al 2011 {\it Mon. Not. R. Astron. Soc.} {\bf414} 3117

\bibitem{30} Demorest P B, Ferdman R D, Gonzalez M E et al 2013 {\it ApJ} {\bf762} 94

\bibitem{31} Hobbs G, Archibald A, Arzoumanian Z et al 2010 {\it Class. Quant. Grav.} {\bf27} 084013

\bibitem{32} www.skatelescope.org/

\bibitem{33} Sesana A and Vecchio A, 2010 {\it Class. Quant. Grav.} {\bf27} 084016

\bibitem{34} Maggiore M, 2008 {\it Gravitational Waves I: Theory and Experiments, Oxford university press} p184-188

\bibitem{35}Sillanpaa A, Takalo L O, and Pursimo T et al 1996 {\it Astron. Astrophys.} {\bf305} L17

\bibitem{36} Lehto H J and Valtonen M J 1997 {\it ApJ} {\bf484} 180

\bibitem{37} Zhang Y, Wu S G, and Zhao W 2013 {\it arXiv:1305.1122}

\bibitem{38} Wahlquist H 1987 {\it Gen. Relativ. Gravit.} {\bf19} 1101

\bibitem{39} Ellis J A, Jenet F A, and McLaughlin M A 2012 {\it ApJ} {\bf753} 96

\bibitem{40} Anholm M, Ballmer S, Creighton J D E et al 2009 {\it Phys. Rev.} D {\bf79} 084030

\bibitem{41} Verbiest J P W, Bailes M, van Straten W et al 2008 {\it ApJ} {\bf679} 675

\bibitem{42} Yardley D R B, Hobbs G B, Jenet F A et al 2010 {\it Mon. Not. R. Astron. Soc.} {\bf407} 669

\bibitem{43} Ma C, Arias E F, Eubanks T M et al 1998 {\it Astron. J} {\bf116} 516


\end{thebibliography}
\end{document}